# Heterogeneous Demand Effects of Recommendation Strategies in a Mobile Application: Evidence from Econometric Models and Machine-Learning Instruments

**Panagiotis Adamopoulos**
Emory University
padamop@emory.edu

**Anindya Ghose**
New York University
aghose@stern.nyu.edu

**Alexander Tuzhilin**
New York University
atuzhili@stern.nyu.edu

## Abstract

In this paper, we examine the effectiveness of various recommendation strategies in the mobile channel and their impact on consumers' utility and demand levels for individual products. We find significant differences in effectiveness among various recommendation strategies. Interestingly, recommendation strategies that directly embed social proofs for the recommended alternatives outperform other recommendations. Besides, recommendation strategies combining social proofs with higher levels of induced awareness due to the prescribed temporal diversity have an even stronger effect on the mobile channel. In addition, we examine the heterogeneity of the demand effect across items, users, and contextual settings, further verifying empirically the aforementioned information and persuasion mechanisms and generating rich insights. We also facilitate the estimation of causal effects in the presence of endogeneity using machine-learning methods. Specifically, we develop novel econometric instruments that capture product differentiation (isolation) based on deep-learning models of user-generated reviews. Our empirical findings extend the current knowledge regarding the heterogeneous impact of recommender systems, reconcile contradictory prior results in the related literature, and have significant business implications.

# Heterogeneous Demand Effects of Recommendation Strategies in a Mobile Application: Evidence from Econometric Models and Machine-Learning Instruments

## 1. Introduction

Mobile devices have become a major platform for information as consumers spend an increasing amount of time with them and use them to search for various products. Mobile devices have also been driving the fast growth in e-commerce sales whereas, at the same time, the contribution of traditional shopping channels has been declining. These trends are mainly attributed to smartphone users and are expected to continue for several years as the number of mobile users has already surpassed the desktop users (StatCounter Global Stats 2016). Projections show that the number of mobile shoppers will soon reach 223.7 million, with 83% of smartphone users shopping online using their mobile devices (eMarketer 2019), while the number of search queries will almost double and the amount of sales will triple (UBS 2015).

At the same time, despite the widespread penetration of mobile devices and the recent advances of technology, information-overload problems and search costs are acuter in such platforms, compared to desktops, due to various technical characteristics and idiosyncrasies of mobile devices. These unique characteristics include, among others, the smaller screen size of mobile devices, the distinct human-computer interaction, and the increased impact of the external environment (Ghose 2017).

Recommender system (RS) techniques, however, offer the potential to further increase the usability of mobile devices and alleviate some of the implications of the aforementioned idiosyncrasies by providing more focused and accessible content and effectively limiting the adverse effects of information overload. Nevertheless, prior research has not identified so far a significant effect of RSs on demand in the mobile channel while various recommendation strategies might have different effects in the mobile world compared to other settings, due to the aforementioned idiosyncrasies and the differences in consumer behavior in the mobile channel. Hence, better understanding and measuring the effectiveness of various recommendation strategies in the mobile context is of paramount importance, given the significance of mobile platforms and the emerging opportunities of RSs. Similarly, to fully leverage any benefits of RSs in the mobile channel, it is also essential to understand any differences in the effectiveness of recommendations across items, users, and recommendation contexts in the mobile channel.

However, there has been scant academic research regarding the economic impact of RSs, especially in the context of mobile recommendations. This is primarily because of the inherent difficulty in measuring the economic impact of RSs, the limited availability of appropriate datasets, and the increasingly important privacy concerns RSs raise (Adamopoulos and Tuzhilin 2015). Therefore, even though the impact of RSs on user behavior and economic demand is a promising field of research, our understanding of how various recommendation strategies in the mobile context may affect product demand levels is limited. For instance, 78% of marketers cite lack of such knowledge regarding RSs as a barrier to their adoption in mobile settings as well as to successful implementations of marketing strategies (Econsultancy.com 2013; Henrion 2019), while mobile RSs have been adopted by only 13% of the companies across the globe and 83% of marketers consider this their biggest challenge (Econsultancy.com 2013; Monetate 2019). A review of the literature indeed confirms that little work has been done on the economic effects of mobile RSs (Li and Karahanna 2015). The current study aims to fill this important gap and examine the demand effects of various mobile RS strategies across different items and contexts.

In particular, part of the first contribution of this study is that we *examine the heterogeneous effectiveness and economic impact of various real-world recommendation strategies in the mobile channel* measuring the increase in *demand for individual items across various factors and settings*. More specifically, we estimate the heterogeneous impact of multiple types of mobile recommendations on consumers' utility and real-world demand for (individual) products, based on a field study with data from a popular mobile application, using an *econometric structural method* that follows the long history of discrete-choice models for product-demand estimation (e.g., (McFadden 1980)). As part of this contribution, we delve further into the differences in the effectiveness of mobile RS strategies and examine the *heterogeneity* of the demand effect, investigating the moderating effect of various item and user attributes as well as contextual factors on the effectiveness of mobile RSs. We theoretically integrate our research questions and findings into the current literature on RSs by extending a conceptual model of the effects of RS use, RS characteristics, and other factors on consumer decision making. This study also makes a secondary methodological contribution. In particular, we *facilitate the estimation of causal effects in the presence of endogeneity* using machine-learning methods. More specifically, we develop novel econometric instruments based on machine-learning models of user-generated reviews. The developed econometric instruments extend the family of the well-established BLP-style instruments of product "isolation" and differentiation (Berry et al. 1995) from the observable product-characteristics space to the latent (product-characteristics) space, and they are applicable in various settings and applications.

Our results show *noteworthy differences* in terms of economic effectiveness and demand effects *in the mobile setting across various popular recommendation strategies incorporating different mechanisms*. Interestingly, recommendation strategies that directly embed *social proofs* for the recommended alternatives (e.g., 'quality', 'expert', and 'trending' recommendations) outperform other recommendations (e.g., 'event' recommendations). Besides, recommendation strategies combining social proofs with higher levels of *induced awareness* due to the prescribed temporal diversity (e.g., 'trending' recommendations) have an even stronger effect on the mobile channel, compared to other popular recommendation strategies based on historical trends and data in the mobile world, highlighting the importance of "in-the-moment" information on the mobile world. The effects are both statistically and economically significant and can have a greater impact on demand than specific item attributes; a 10% increase in the number of times a product (alternative) is recommended raises demand by about 7.2% on average for the corresponding alternative, increasing aggregate demand too, while the effect is largely heterogeneous across strategies. We theoretically explain the empirical findings and uncover the underlying mechanisms drawing from the persuasion theory and social cognitive decision-making theories, including symbolic interaction theory and resource matching theory. Notably, we also find that *not only various item and user attributes but also contextual factors significantly moderate the effectiveness of recommendations*, further illustrating the heterogeneity of the demand effect of mobile RSs while generating rich insights and further verifying the identified mechanisms. The findings and mechanisms are robust to incorporating and measuring heterogeneity across multiple other factors, such as markets, users, and different RS implementations. Our empirical findings *extend the current knowledge* regarding the impact of RSs filling an important gap in the current literature, *reconcile contradictory prior results* in the related literature, and draw significant *business implications*.

## 2. Literature Review and Research Questions

Our work is related to several streams of research. In the next paragraphs, we focus on the most relevant works while also discussing how this study extends the conceptual model of RSs effects (Li and Karahanna 2015; Xiao and Benbasat 2014); Figure A1 depicts the conceptual model incorporating our research questions, with all new constructs depicted in red font and italics.

Prior literature has examined specific effects of desktop recommender systems focusing on *sales diversity* and total *aggregate market* demand. Examining such consumption patterns, De et al. (2010) and Dias et al. (2008) find that desktop RSs are associated with an increase in total aggregate sales. Moreover, focusing on sales diversity, Fleder and Hosanagar (2009) analytically show that RSs can lead to a reduction in aggregate sales diversity. However, Brynjolfsson et al. (2011) and Hinz et al. (2011) provide empirical evidence that RSs are instead associated with an increase in aggregate sales diversity, reflecting lower search costs in addition to the increased product availability, corroborating the findings of Pathak et al. (2010). Nevertheless, studying whether RSs are fragmenting the online population, Hosanagar et al. (2014) find that users widen their interests, which in turn creates commonality with others instead of heterogenizing users. In this study, however, we focus on demand levels for individual products especially on RSs in the mobile context and the heterogeneity of these effects across different recommendation strategies, rather than impact on sales diversity and aggregate demand at the market level that the aforementioned stream of work has examined for desktop RSs.

Even though the majority of prior studies have focused on the effects of desktop RSs on concentration bias and the aggregate market level or networks of hyperlinked products in non-mobile contexts (e.g., (Oestreicher-Singer and Sundararajan 2012; Pathak et al. 2010)), preliminary computer science case studies examine the effect of recommendations on demand levels for individual products in mobile platforms. Jannach and Hegelich (2009) present a case study evaluating the effectiveness of item recommendations for mobile apps (combined paid and free apps) and find that "recommender systems did not measurably help to turn more visitors into buyers." However, this case study neither separates paid from free apps nor considers the app prices, as consumers' utility, willingness-to-pay, and economic demand are all beyond the scope of their study. Besides, prior work has focused on collaborative filtering RSs, even though the alternative paradigms of content-based and hybrid RSs entail different characteristics (Adomavicius and Tuzhilin 2005; Xiao and Benbasat 2014) prescribing different mechanisms, as described in the next section. In this study, we aim to fill these important gaps in the literature and, therefore, first focus on the assessment of the –*potentially heterogeneous*– *impact of various real-world mobile RS strategies on demand levels for individual products* and the utility of consumers in the context of the mobile channel. More specifically, using actual data corresponding to all the users and alternatives of a popular real-world mobile application, we first examine the following research question:

- **RQ 1**: What is the economic effect of various RS strategies on real-world demand for individual items in the mobile context, and is it heterogeneous across different RS strategies?

This will allow us to unveil the effects of the various RS strategies on the demand levels for individual products in the mobile channel based on a rigorous structural demand-estimation model for differentiated products. In order to provide additional insights for the underlying mechanisms, we further examine the heterogeneity of these effects in order to gain a more detailed understanding of the effectiveness of the various recommendation strategies in the mobile setting and uncover the corresponding mechanisms. In particular, we examine whether various *contextual factors* (e.g., *traffic, mobile network performance)*, as well as *item attributes* (e.g., *price*) and certain *user traits* (e.g., *income*) moderate the effectiveness of different types of mobile recommendations. More specifically, we examine the following research question:

- **RQ 2**: Is the economic effect of RS strategies on real-world demand in the mobile context heterogeneous across different contextual factors, item attributes, and user traits?

Despite the importance of context in the mobile channel, no prior work has examined the moderating effect of *contextual factors* on the effect of mobile RSs on demand levels. Similarly, no prior work has examined the moderating effect of specific item attributes. Such attributes can deepen our understanding of the effectiveness of RS strategies, including whether alternatives with higher or lower profit *margins* are benefiting from recommendations. Nevertheless, prior research on desktop RSs has examined other specific item attributes as moderators, such as product type, product complexity, and product novelty for this different channel. In particular, Senecal and Nantel (2004) examine the moderating effect of product type and find recommendations are more useful for experience –compared to search– goods. Fasolo et al. (2005) examine the impact of product complexity and find that consumers using desktop RSs engage in more information search and are less confident in their product choices for higher product complexity. Finally, Ekstrand et al. (2014) and Matt et al. (2014) find that novelty in desktop RSs has a significant negative effect on consumers' satisfaction and perceived enjoyment, whereas Vargas and Castells (2011) and Pathak et al. (2010) argue that novelty is a key positive quality of desktop recommendations. Prior work on desktop RSs has also examined specific user attributes, different from those studied here. In particular, Kramer (2007), Swaminathan (2003), and Chang and Chin (2010) examine the effects of product expertise, perceived product risk, and gender, respectively. As described above, no prior work has examined the moderating effect of contextual factors and other user and item attributes on the effect of mobile (or desktop) RSs on demand levels, while there is contradictory evidence for specific item attributes in desktop RSs and calls "for additional studies to resolve inconsistent findings on RS impacts" (Li and Karahanna 2015). These research gaps are evident by *theoretically integrating our research questions into the current literature on RSs* and extending the conceptual model of the effects of desktop RS on consumer decision making that was first articulated by Xiao and Benbasat (2007). Figure A1 depicts the *extended conceptual model*. In particular, *RQ 1* is depicted in the 'RS principle of function' factor that was added to the model, *RQ 2* is depicted in the new 'Context of use' construct (e.g., 'Mobile network), the 'Product' construct (e.g., 'Product price'), and the 'User' construct (e.g., 'Income').

This study also relates to prior work discussing the distinguishing characteristics of the information environment in mobile devices and, in particular, the differences between the characteristics of mobile and desktop devices. These differences include the distinct human-computer interaction, the increased physical and cognitive effort required, the increased impact of the external environment, the variability in the context of usage and ubiquity, etc. (Andrews et al. 2016; Ghose 2017; Ghose et al. 2013). These fundamental differences in the information environments across channels manifest in users' decision-making processes and information perceptions and, hence, affect their shopping decisions and overall consumer behavior. Prior research has shown, for instance, that mobile devices are perceived as more personal, emotional, sensational, and experiential (Jung et al. 2019). As a result, product comparison is more difficult and consumption patterns are significantly different (Ghose and Han 2011; Huang et al. 2016).

Finally, our work is also related to the stream of literature that integrates machine learning and data mining with econometric techniques. In particular, in this study, we employ deep-learning-related methods drawing theoretical support from the linguistic theories of Harris (1954) and Osgood et al. (1957) in order to introduce new machine-learning-based econometric instruments that extend the popular family of BLP instruments (Berry et al. 1995) from the observed product-characteristics space to the latent space. Such a machine-learning-based approach has the potential to facilitate the estimation of causal effects in the presence of endogeneity leveraging the abundance of unstructured data, such as user-generated content, when structured product attributes are either not available or not sufficient or endogenous. Hence, our work is related to the extant literature in Information Systems that employs text mining, sentiment analysis, and other data-mining methods with user-generated content in econometric studies (e.g., (Adamopoulos 2013b; Adamopoulos et al. 2018; Adamopoulos and Todri 2015; Adamopoulos et al. 2020; Archak et al. 2011; Kokkodis 2021)). Prior studies have employed data-mining and machine-learning techniques to automate the measurement of dependent and independent variables, facilitating observational studies to be conducted with larger samples. This work, however, moves beyond facilitating larger samples and enables causal inference by developing econometric instruments based on machine-learning methods that capture product isolation (differentiation) extending a seminal family of instruments to the latent product-characteristics space. Much as a randomized trial obviates extensive controls in regression, the proposed approach enables causal inference even in cases of omitted or unknowingly missing variables, which is not feasible with several other causal inference methods (Angrist and Pischke 2008). Thus, the proposed approach and the developed instruments can facilitate causal inference in the presence of any endogeneity concerns in several fields, such as RSs and demand estimation.

## 3. Mobile Recommendations and Data

We examine our research questions employing a large data set from a very popular real-world mobile platform that identifies and recommends interesting venues to users. In aggregate, our dataset includes 12,119 venues and the corresponding (physical) visits to these venues of several tens of million active users from February 2015 until March 2015. In particular, our data set includes all the restaurants in the mobile urban guide app for the ten most popular cities (in terms of population) in the United States.

More specifically, the dependent variable (DV) of our econometric analyses corresponds to the total number of visits to a particular venue in a single time period (i.e., day). The independent variables (IVs) of interest in our analyses include various recommendation strategies, such as recommendations based on past *historical trends and data*. In particular, the different types of alternative (venue) recommendations include recommendations based on whether a venue is recommended because of the total number of submitted positive *user-generated reviews*, positive *ratings* from the users, etc. (i.e., 'quality-based recommendations'), as well as recommendations based on whether a business is endorsed through reviews by *famous brands and experts* (i.e., 'expert recommendations'). Both these recommendation strategies belong to the standard paradigm of collaborative filtering (Adomavicius and Tuzhilin 2005; Gorgoglione et al. 2019).[1] A distinguishing characteristic of these recommendation strategies is that they are prescribed with information on what others are doing (Xiao and Benbasat 2014) and, hence, directly embed a *social proof for the recommended alternatives* (Cialdini 2007). Such social proof is prescribed in the corresponding strategies in the form of peers' implicit and explicit ratings, experts' reviews, etc. The IVs include recommendations for *novel* venues too, as alternatives that opened recently are also recommended to the users (i.e., 'novel recommendations'). This recommendation strategy is prescribed with further enhancing awareness focusing on novel alternatives (Adamopoulos 2014; Adamopoulos and Tuzhilin 2014a), instead of embedding social proofs. In addition, the mobile application also recommends to the users venues that have scheduled *upcoming events* for customers (i.e., 'event recommendations'); hence, this type of recommendations does not embed any of the aforementioned mechanisms. These two latter recommendation types belong to the content-based paradigm. The IVs also include whether a venue is recommended as a "*trending*" venue (i.e., 'trending recommendations'). This type of hybrid recommendations takes into consideration the latest trends and data for the specific alternatives, by explicitly discounting past historical data while leveraging available information that captures current trends based on the normalized relative differences in item attributes (e.g., number of photos) during the last time periods, so that interesting alternatives –and not only the most popular ones– are recommended to the users (Adamopoulos 2013a; Yang and Sklar 2016). This type of recommendations is designed to take advantage of the higher involvement of consumers with mobile devices and to leverage the differences in consumption patterns (e.g., more instantaneous and less planned behaviors) in mobile platforms, capturing trends related to "*in-the-moment" marketing*. In addition to embedding social proofs, this recommendation strategy also *enhances awareness* to a larger extent by design focusing more on recent patterns and discounting past historical data, as previously described (Gorgoglione et al. 2019). Our dataset includes all the recommendations implemented in the platform and contains –for each type of recommendations and (time) period– the relative number of times a venue was recommended to users, as well as the ranking of the venue in the generated recommendation lists. Apart from the distinction between different recommendation strategies (i.e., 'quality', 'expert', 'novel', 'event', and 'trending' recommendations), all the recommendations are seemingly similar as they are presented to the users in the same way. Finally, the relevant IVs include for all the alternatives in our dataset how many brands and experts have endorsed each venue and how many upcoming events are scheduled.

---

[1] Content-based approaches have their roots in information retrieval and information filtering research and use information on item characteristics, whereas collaborative filtering approaches have their roots in stereotyping and use information of other users, while hybrid approaches combine collaborative, and content-based methods using all inputs of the other approaches (Adomavicius and Tuzhilin 2005; Gorgoglione et al. 2019).

The aforementioned recommendation strategies entail several advantages for the identification of relevant empirical models. For instance, all the recommendations are generated before the realization of the corresponding demand for each alternative; all recommendations are pre-computed offline during the night while the alternatives are not available to consumers. Besides, all the recommendation lists are explicitly diversified by the RS algorithms (Ziegler et al. 2005), enhancing the exogeneity of recommendations. In other words, the diversification process provides an *exogenous shock* to the recommendation process, further facilitating our identification strategy. In particular, the mobile app uses a top-down diversification approach – based on a greedy optimization algorithm– to re-rank the initial recommendation list in order to increase the diversity (i.e., differentiation) of the alternatives in the list, explicitly taking into account how different each alternative is compared to the alternatives that are already included in the list. Beyond further enhancing variation within and between alternatives providing an exogenous shock, the diversification process alters the recommendations without directly affecting the alternative demand levels and, hence, can also help us develop appropriate econometric instrumental variables, as discussed in the following sections.

Additionally, our dataset includes a large number of item attributes to control for additional confounders. In particular, our data set includes the number of photos uploaded by users for this venue, the average numerical rating of the venue and the corresponding number of ratings, the price tier of the venue (i.e., from 1 to 4 with 1 corresponding to least pricey), whether the venue is part of a chain, the exact location of the venue, the various categories users have applied to this venue (e.g., American restaurant, Vegetarian, etc.), whether a marketing promotion (e.g., discount) is taking place during that time period, whether the venue offers breakfast, brunch, lunch, dinner, alcohol, delivery, or take-out, whether the venue allows reservations, whether it accepts credit cards, whether it has live music, DJs, TVs, Wi-Fi, outdoor seating, and parking availability, whether it is wheelchair accessible, when the venue opened and was first introduced in the platform, as well as a description of the venue and all the user-generated reviews posted on the mobile app. Hence, beyond the advantages mentioned above, our dataset allows us to observe, among other effects, marketing promotions, user- and expert-generated content, and real-world economic demand −not just digital clicks or similar user actions− in a domain that is appealing to a large portion of the population (e.g., (Kokkodis and Lappas 2020)).

### 3.2 External Data Sources

We further supplement our dataset with additional *contextual variables* from external sources. In particular, for each of the venues in our data set, we include *climate data* (e.g., temperature and precipitation) from the National Oceanic and Atmospheric Administration, *calendar data* with information about holidays and weekends, *mobile network performance* data from Root Wireless, Inc., and *traffic levels* based on open data from the Divvy Bikes sharing system. Such contextual factors will allow us to examine the potential moderating effects of context in our online-to-offline setting. In addition, beyond these contextual variables, we include additional attributes from external sources, including *rental prices* (e.g., median rental price at each location) from the Zillow Group and demographics from the American Community Survey of the *US Census* Bureau, as discussed in the next sections. Finally, as part of the robustness checks, *additional external data* sources are incorporated in our dataset, including transit information as well as external sources of information for users, as described in Section 7.3. Table 1 contains summary statistics that describe the variables of the main interest.

## 4. Empirical Method and Structural Models

We next discuss the econometric demand-estimation structural model we apply to estimate the effect of the various RS strategies based on the utility of consumers regarding the different alternatives and their sensitivity to changes in these utility components. In a nutshell, each consumer selects the alternative that gives her/him the highest utility, while the utility of consumers depends on the alternative characteristics, specific contextual factors, whether the particular alternative is recommended by the mobile application, as well as individual taste parameters. The alternative (market) shares are then derived as the aggregate outcome of individual consumer decisions and the utility parameters are inferred based on these decisions. In the following paragraphs, we explain in detail the econometric structural models we apply in order to estimate models of individual behavior with aggregate data and unobservable by the econometrician effects; Table 2 summarizes the employed mathematical notation.

In particular, there are $R$ markets (i.e., cities) with $N_r$ alternatives (i.e., venues) in market $r$. For each alternative $j$ in market $r$ and time period (i.e., day) $t$, the observed characteristics are denoted by vector $z_{jrt} \in \mathbb{R}^{K_z}$, contextual factors by vector $w_{jrt} \in \mathbb{R}^{K_w}$, and recommendation types by vector $\rho_{jrt} \in [0,1]^{K_\rho}$; for simplicity, $z_j$, $w_j$, and $\rho_j$, respectively. The elements of $z_j$, $w_j$, and $\rho_j$ combined include observed attributes $x_j$ (e.g., quality, frequency of each mobile recommendation strategy, etc.) that affect the demand levels $q_{jrt}$ (i.e., number of visitors); for simplicity, $q_j$. The unobserved characteristics (e.g., perception of status or aesthetics) of alternative $j$ are denoted by $\xi_j$. The utility $u_{ij}$ of user $i$ for alternative $j$ depends on the characteristics of the alternative and the user as well as the price $p_j$ and recommendations $\rho_j$.

In addition to the competing alternatives $j = 1, ..., N$, we model the existence of "the outside option", $j = 0$. This outside option corresponds to alternatives that might not be present in our data set or the option of a user not choosing any alternative at all in period $t$. Consumers may choose to select the outside option instead of the $N$ "inside" alternatives; the mean utility value of the outside option is normalized to zero. Following the standard assumptions for utility maximization (i.e., the consumer chooses the alternative that maximizes his/her utility surplus) and assuming that $\epsilon_{ij}$, which captures user-specific taste parameters, follows an extreme value distribution[2], the probability that a user $i$ chooses alternative $j$ is (McFadden 1980):

$$\Pr\left(\text{choice}_j^i\right) = \frac{e^{u_{ij}}}{\sum_{k=0}^N e^{u_{ik}}} = \frac{e^{\beta x_j - \alpha p_j + \xi_j}}{1 + \sum_{k=1}^N e^{\beta x_k - \alpha p_k + \xi_k}}, \qquad (1)$$

$\forall\, k$ in the same market $r$ and $k \neq j$, while $\beta$ and $\alpha$ represent the taste of the consumers regarding the corresponding observed attributes.

The market share $s_{jrt}$, for simplicity $s_j$, of each alternative is then calculated as $s_j = q_j/M_r$, where $M_r$ is the total market size for the corresponding city (i.e., market) $r$. This market size $M_r$ is set to the maximum number of unique active users that has been ever observed in the mobile application for that specific city. Alternatively, the market size could be assumed to be the population of each city or the number of households; the results remain qualitatively the same. Inverting the market-share equation and taking the logarithm in Eq. (1), the market share of alternative $j$ is (Berry 1994):

$$ln(s_j) - ln(s_0) = \beta x_j - \alpha p_j + \xi_j, \qquad (2)$$

where $s_0$ represents the market share of the outside option. This specification corresponds to a logit discrete-choice model of individual behavior that can be estimated with aggregate data and relates individual utility levels to alternative characteristics, recommendations, and other factors.

---

[2] The distribution of $\epsilon$ can be interpreted as representing the effect of factors that are quixotic to the consumer himself (representing, e.g., aspects of bounded rationality) (Train 2009).

Additionally, if we assume user tastes are correlated across alternatives and group (nest) the alternatives into $G$ exhaustive and mutually exclusive sets of similar alternatives, $g = 1, \dots, G$, the market share of alternative $j$ is (Cardell 1997):

$$ln(s_j) - ln(s_0) = \beta x_j - \alpha p_j + \sigma ln(\bar{s}_{j/g}) + \xi_j, \qquad (3)$$

where $\bar{s}_{j/g}$ is the market share of alternative $j$ as a fraction of the total group (nest) share and $j$ is in group $g$. As the parameter $\sigma$ approaches one, the within-group correlation of utility levels goes to one, and as $\sigma$ approaches zero, the within-group correlation goes to zero. This specification corresponds to a nested-logit discrete-choice model of individual behavior that can be estimated with aggregate data.

In other words, using demand-estimation approaches for differentiated products from economics, we estimate the weights consumers (implicitly) assign to alternative characteristics, mobile recommendations, and contextual factors. This is done by inverting the function defining market shares to uncover the utility levels of the alternatives and relating these utility levels to alternative characteristics, recommendations, and contextual factors. Then, based on these estimates, we derive the utility gain each RS strategy generates.

Apart from the dataset-related benefits (e.g., popular real-world application, observable economic demand, quantifiable quality, observable marketing promotions, exogenous variation, various mobile RS strategies, multiple markets, etc.), this particular modeling method allows for unobserved product characteristics, including determinants that are difficult to measure (e.g., consumers' perceptions about status) (Crawford 2012). Similarly, apart from unobserved product characteristics, this method allows for unobserved portions of each consumer's utility and idiosyncratic preferences. Besides, the model does not require normalizing the mean utility of one of the inside options while allowing the consumers the option of not choosing any alternative. Additionally, the resulting empirical model can make predictions not only for the existing alternatives under different conditions and contextual factors but also for new alternatives not included in our data. Hence, not only does the employed structural model allow for the estimation of an individual behavior model when only aggregate data are observed with unobservable effects, but the findings are also more generalizable compared to other approaches that simply model the choice of consumers among alternatives conditional on their choosing one of the alternatives (Train 2009). These strengths of the model are coupled with several widely established methodologies for causal inference, as described in the next paragraphs.

### 4.2 Identification Strategy

We can treat Eq. (2) and (3) as estimation equations, treating $\xi_j$ as an unobserved error term, and use typical econometric techniques to estimate the parameters of our structural models capturing the effects of the variables of interest. In particular, we employ panel-data techniques to further control for unobserved confounders, while we leverage the within-alternative variation, the explicit diversification of the recommendations by the mobile RS, and the methodological benefits described above.

### 4.2.2 Instrumental Variables

Nevertheless, we allow for the possibility that the variables of interest in our econometric specifications are endogenous. Hence, we use novel econometric instruments derived from a metric of (alternative) differentiation and isolation based on machine-learning models of the user-generated reviews employing deep learning as well as specific variables that are used in the algorithms to generate the recommendations but do not affect the utility of the alternatives for consumers in the current time period given the observed confounders. The motivation for the novel econometric instruments is that they are related to whether an alternative is recommended by the mobile application as alternatives that are more isolated in the product space have a higher likelihood of being included in a recommendation list because of the diversification process of the employed RS while they are uncorrelated with the $\xi_{jrt}$ term given the covariates. One significant benefit of the developed machine-learning-based econometric instruments is that, as discussed in the following section, they extend the popular family of BLP-style instruments (Berry et al. 1995; Berry 1994) from the observable space of product characteristics to the latent space, better capturing the corresponding relationships; intuitively, the level of alternative isolation (differentiation) in the latent product space reflected by the developed instruments corresponds to spatial locations in a Hotelling-like space. Hence, the developed instruments can facilitate causal inference in the presence of endogeneity concerns even when structured product attributes are not available or not sufficient or endogenous (see Sections 7.1 and 7.2 for applications). Additional theoretical justification and motivation for these instruments, technical details, statistical tests, and additional applications are provided in the next sections.

## 5. Machine-Learning Models of User-Generated Reviews
### 5.1 Overview of Developed Machine-Learning Instruments for Econometric Models
To implement the aforementioned machine-learning-based instruments that measure the differentiation of alternatives in the latent space of products, we use an efficient and state-of-the-art method based on deep-learning techniques (Le and Mikolov 2014). We use these techniques to generate based on textual reviews (continuous distributed) latent space representations –also known as "embeddings"– of the alternatives that are then used in our method to measure the level of alternative differentiation (isolation) in the latent space and develop econometric instrumental variables for causal inference, extending the BLP instruments to the latent space.

The prevailing paradigm for deriving such latent-space representations from variable-length user-generated text is based on the vector intuition of Osgood et al. (1957) and the distributional hypothesis of Harris (1954), theorizing that the meaning of words can be modeled as points in a multidimensional latent space and can be inferred from text as words in similar contexts have similar meanings (Levy and Goldberg 2014). In particular, latent space vector representations of texts representing various entities (e.g., products, users, etc.) are learned by training a machine learning model to simultaneously learn (distributed) vector representations of words and the entities of interest by predicting whether a word will be found in a given text context (e.g., user phrase in a review of the corresponding entity) using the entity and word representations as features (e.g., the 'Paragraph Vector' algorithm (Le and Mikolov 2014)). Both the alternative (entity) and word dense numeric vector representations, learned as part of the prediction task, reflect similarities and differences among words and entities (Le and Mikolov 2014).[3]

These continuous latent space representations of the alternatives that are learned based on the above deep-learning procedure can then be used to capture the differentiation (isolation) of alternatives in the latent space utilizing an angular distance metric.

### 5.2 Additional Implementation Details and Novelties

In our empirical setting, in order to learn the numeric representations of the alternatives in the latent space and then estimate the corresponding distance (isolation) of the various alternatives based on an angular distance metric, we consider all the available user-generated reviews of each alternative (venue) at each time period as a single document of our corpus (input)[4,5] and use a pre-trained open-source neural network model that contains 300-dimensional vectors (word representations) for 3 million words and phrases trained on the Google News dataset (about 100 billion words -  https://code.google.com/archive/ p/word2vec/). In essence, this open-source neural network model provides weights for the word layer in our application further enhancing the reproducibility of this approach (the weights for the entity layer are inferred as part of the training); the findings remain robust to alternative implementations.

---

[3] Representing words and phrases as dense numeric vectors is one of the advantages of neural embeddings compared to other text-mining representations, such as bag-of-words or TF-IDF, that not only work in terms of discrete units without meaning that have no inherent relationship to one another, but also suffer from data sparsity and high dimensionality (Le and Mikolov 2014). Bag-of-words models, for instance, are not as informative because every two distinct vectors are the same distance from each other (Goodfellow et al. 2016).

[4] The input (corpus) corresponds to all the available user-generated reviews at the time of generating the recommendations. Hence, the input only includes user-generated reviews written before $\xi$ is known.

[5] As the user-generated content is expanded over time, the generated econometric instruments exhibit both cross-alternative and within-alternative variation over time. Such variation is further enhanced through changes in the set of available alternatives in each market.

The neural network models for these embeddings contain $(J + V) \times q$ neurons for the word and entity layers of the model, where $J$ is the number of entities (i.e., alternatives here) in the entity layer (matrix $D$), $V$ the size of the vocabulary (i.e., number of words) in the word layer (matrix $W$), and $q$ the number of latent dimensions in the representation, and can be trained on corpora (e.g., user-generated reviews) that contain billions of words, resulting in high-quality numeric representations (Mikolov et al. 2015);[6] see (Rong 2014) for a detailed step-by-step description of the learning phase of such neural networks and the 'Paragraph Vector' algorithm (Le and Mikolov 2014) for the exact architecture of this particular neural network we employed with the multidimensional hidden layer containing the entities and word matrices. The neural-network-based continuous embeddings for the entities are unique among entities in our corpus while the word embeddings are shared across entities.

Then, we use the cosine angular distance to measure the isolation (differentiation) of each alternative in each market based on the entity (alternative) numeric representations (neural embeddings). Compared to the angular distance, other commonly used functions, such as the standard cosine distance, do not have the Cauchy-Schwarz inequality property. This is a significant advantage for the construction of the econometric instrumental variables as we are interested in the distance measures of all alternatives and not only the relative ranking of distances. Besides, the angular distance better distinguishes similar representations compared to the cosine distance, for instance.

---

[6] The number of latent dimensions in the representation of entities can be different from the number of latent dimensions in the representation of words.

We should also note that, compared to other differentiation metrics, the developed instruments do not depend on endogenous market shares (e.g., (Tallman and Li 1996)) or external information, such as financial statements or industry classifications (e.g., (Varadarajan 1986)) or other domain-specific information (e.g., (Nguyen et al. 2018)). Besides, following the proposed approach, we avoid overfitting and inducing any violation of the independence of treatment without sacrificing efficiency. The validity of the instruments and their robustness to various design choices is further verified in Section 7, where we discuss their broader applicability.

In summary, we propose a novel application by generating econometric instrumental variables that extend BLP-style instruments to the latent space using an angular distance and NLP neural networks. Utilizing neural representations that achieve a level of generalization that is not possible with classical n-gram language models allows us to capture the alternative differentiation in the latent space and, hence, to develop novel instrumental variables for causal identification of structural demand-estimation models, illustrating the complementarity of the corresponding machine learning and econometric methods and creating a synergy that significantly facilitates causal inference. Beyond the theoretical justification and motivation for the developed instrumental variables, the developed instruments are valid as indicated by multiple tests and statistics and remain robust to various design choices (see Section 7.1), while they are also applicable in additional applications and settings (see Section 7.2). Even though we illustrate the proposed approach utilizing textual data, in principle, any deep-learning method for generating entity representations in the latent space can be employed regardless of the nature of the input (e.g., images).

## 6. Empirical Results

To discover the impact of the different recommendation strategies in the mobile world, we estimate different specifications of the structural econometric models we presented in Section 4. *Interestingly, we find statistically and economically significant differences in effectiveness among the different mobile recommendation strategies*. We *further examine the heterogeneity of the demand effects estimating several interaction effects based on contextual factors, item attributes, and user characteristics and verify the identified mechanisms*. The empirical findings extend the current knowledge regarding the impact of RSs, reconcile contradictory prior results in the literature, and draw significant managerial implications.

### 6.1 Overview of the Effectiveness of Different Recommendation Strategies

In particular, Table 3 shows the coefficient estimates for alternative demand and the corresponding impact of the various recommendation strategies in the mobile context, based on the nested logit model with multi-level fixed effects (i.e., market, venue category), various alternative and context controls (e.g., number of events, meals served, location, holiday, temperature, precipitation), and a linear time trend as well as day-of-the-week effects, in order to control for correlation of tastes across alternatives as well as different potentially unobserved effects. Model 1 identifies the average effect of recommendations in general, using the frequency of recommendation of the specific alternative. Then, Model 2 separates the effects of the different recommendation strategies (i.e., 'quality recommendations', 'event recommendations', 'expert recommendations', 'novel recommendations', and 'trending recommendations'). Finally, Model 3 controls for the ranking of each alternative in the generated recommendation lists. Due to space limitations, only coefficients of the main variables of interest and statistically significant effects are shown in the corresponding tables; all the controls are included in all the employed specifications and conducted analyses.

(Insert Table 3 about here)

As Models 1-3 in Table 3 show, *recommendations have a positive impact on demand* levels for individual alternatives (products) in the mobile recommendation setting. Interestingly, there are *statically and economically significant differences across the various recommendation strategies*. Based on Model 3 in Table 3, the coefficients for the various mobile recommendation strategies range from 1.4908 for 'trending' recommendations to 0.3399 for 'novel' recommendations; this difference is statistically significant and corresponds to a lift of 3.39. These interesting findings illustrate the ***heterogeneity*** *of the demand effects across the different mobile recommendation strategies* (RQ1). This statistically significant effect heterogeneity persists even after controlling for the ranking of each alternative, as shown by Model 3.

These findings are especially interesting as the recommendation strategies of 'trending', 'quality', and 'experts', which directly embed a ***social proof*** for the recommended alternatives (Cialdini 2007), outperform the strategies of 'novel' and 'events', which do not contain any social proofs. This interesting finding is in accordance with prior literature in social cognition and, in particular, the *persuasion theory* suggesting that social proofs provide a valuable decision heuristic to consumers (Petty and Cacioppo 1986; Sherif 1936). Hence, social proofs embedded in these recommendation strategies effectively constitute a persuasion mechanism for consumers and lead to the observed increase in demand in the mobile world; we further validate this mechanism in the next section. Given the information environment in the mobile world and cognitive decision theory (e.g., (Petty and Cacioppo 1986; Tversky and Kahneman 1974)), consumers undertake peripheral processing invoking available heuristic decision rules and accessible rules of thumb triggering inferences and judgments based on the recommendations (Tam and Ho 2005). Interestingly, mobile RSs embedding social proof are effective persuasion mechanisms even though the choice of a recommended alternative is not directly observable by others, yet this finds support in symbolic interaction theory (Schlenker 1980). Whereas, 'event' recommendations are a less effective strategy based on the results. This is in accordance with the persuasion theory mechanism, as the elaboration likelihood of consumers is lower due to the relatively lower relevance of future events, according to the resource matching theory (Anand and Sternthal 1989).

In addition, it is also interesting to note that the 'trending' recommendations, which combine social proofs higher levels of ***induced awareness*** due to the prescribed temporal diversity, outperform the corresponding methods of 'quality' and 'experts' while, similarly, 'novel' recommendations outperform 'events' for the same reason. In other words, beyond directly persuading consumers, recommendation strategies also increase consumers' awareness and enhance their corresponding consideration sets and, hence, recommendation strategies with increased temporal diversity combining both information and persuasion effects outperform other strategies of recommending items of which consumers are more likely to be already aware; we further validate this mechanism too in the next section. This interesting finding is in accordance with social cognition theory (Bohl and van den Bos 2012) and the theory of advertising in economics (e.g., (Bagwell 2007)) illustrating the presence of both informative and persuasive effects of advertising and highlighting the relative effectiveness of the informative role and its combinations (Ackerberg 2003; Ghose and Todri-Adamopoulos 2016).

We should note here that prior research has not thoroughly examined such heterogeneity of demand effects and, at the same time, the theories used in prior related work –such as the theory of constructed preferences, consumer search theory, social network theories, trust theories, etc.– do not fully explain this heterogeneity. This study fills this important gap in prior literature.

### 6.1.2 Alternative Model Specifications

We examine several additional model specifications enhancing our identification strategy. For instance, Table A1 in the appendix presents the results for the nested logit demand model controlling for unobserved heterogeneity across alternatives by introducing alternative-level fixed effects. The results further substantiate our previous findings regarding the *heterogeneous demand effects across various mobile recommendation strategies*. Note that after controlling for unobserved heterogeneity at the alternative level, the effect of novel recommendations is not found to be statistically significant. This result regarding the novelty of recommendations is interesting in the context of the contradictory results in the extant literature on the effects of desktop RSs; this finding is in accordance with the algorithmic development proposal of Adamopoulos and Tuzhilin (2014b) that novelty should be considered vis-à-vis the overall utility of each alternative when generating recommendations, rather than simply recommending the most novel items. It is also worth noting that the effect of promotional marketing strategies is now positive and significant, indicating in essence that even though marketing promotions have on average a positive effect on demand, such promotions are usually offered by alternatives that experience lower levels of demand than expected.

Moreover, we further extend our heterogeneity analysis by allowing individual deviations at the consumer level, using the BLP algorithm (Berry et al. 1995) with the squared polynomial extrapolation method for fixed-point acceleration (Varadhan and Roland 2008). Table A2 in the appendix presents the corresponding results. The results further corroborate our findings.

Additional specifications accommodating other sources of heterogeneity are presented in the robustness sections; the results remain qualitatively the same.

### 6.1.3 Economic Significance of Recommendation Effects

We further examine the economic significance of the effects of interest computing alternative-level derivatives of the demand function (elasticities). Based on the results, a 10% increase in the number of times an alternative is recommended raises on average the demand by 7.15% for already-recommended alternatives and by 0.92% for all alternatives in general. Hence, *mobile RSs have positive effects on both individual demand and aggregate-level demand* in the market. These effects are *both statistically and economically significant* as well as novel in the context of the extant literature on the mobile channel; prior literature suggests that "recommender systems did not measurably help to turn more visitors into buyers" (Jannach and Hegelich 2009).[7,8] In particular, considering that the average spending on food and drinks per restaurant visit is about $34.63 (Statista 2016), this difference corresponds to an expected demand increase of $3.58 per consumer and day, which sums up to an increase of total sales of about 2.42 billion U.S. dollars (National Restaurant Association 2017).

Even though we have already shown the positive (heterogeneous) effect of recommendation strategies on demand levels, it is also important to examine whether this increased demand is for alternatives with higher or lower profit margins for brands and, hence, we investigate the moderating effect of prices. Based on the results in Table 4, we find a positive and significant moderating effect of price on the effectiveness of recommendations. This finding indicates that

---

[7] The effects of mobile recommendation strategies are larger than the corresponding effects identified in prior literature for desktop RSs, highlighting the differences of consumer behavior in the mobile channel.
[8] Comparing the effect of recommendations in the mobile setting with the effects of various attributes on demand levels, a 1% increase in the frequency of recommendation of an alternative is equivalent to an increase of about 8% in the rating of the alternative, about 4.2% in the number of reviews, and about 13% in the number of photos.

*even though alternatives that are more expensive are less appealing to the users, ceteris paribus,*

*they can more effectively leverage the additional attention they garner from recommendations.*

From a managerial point of view, this further alleviates any potential concerns that mobile recommendation strategies might be pushing users towards items with lower margins, potentially hurting the profitability of firms despite the elevated demand levels. This is especially interesting as prior literature on mobile devices has shown that mobile users, in general, have a tendency to purchase lower-priced alternatives (Huang et al. 2016).

(Insert Table 4 about here)

### 6.2 Underlying Mechanisms and Theoretical Foundations

We further delve into the heterogeneity of the effectiveness of recommendations in the mobile

channel, examining the moderating effects of various item attributes, contextual factors, and user

traits that allow us to assess the discussed mechanisms.[9]

In particular, we first assess the induced awareness mechanism examining the moderating effect

of novelty. If alternatives that are more novel benefit more from recommendations, then this

would further confirm the identified mechanism of induced awareness as consumers are less

aware of such alternatives. Alternatively, if less novel alternatives benefit more, then this would

nullify this identified mechanism. Based on the results presented in Table 5, novel alternatives

gain greater benefits from recommendations. Hence, this analysis further validates the

informative role of recommendation strategies with higher temporal diversity. Besides, in

combination with the previous findings, this result illustrates that recommendation strategies based on just the novelty of the alternatives do not have a significant effect, but *novel alternatives accrue greater benefits from recommendations when those recommendations* are not based solely on the characteristic of novelty but also *take into consideration attributes such as the quality of the item*. This is also highlighted by the magnitude of the effect in the case of recommendations based on quality. Hence, this finding both verifies the identified mechanisms and reconciles the aforementioned seemingly contradictory findings in prior literature. The results are very similar when examining the moderating effect of the *number of alternatives* in the same market and category. These results further verify the identified mechanism of the informative role, as awareness levels for each alternative are associated with the number of available alternatives.

---

[9] All the presented results extend the results presented in Table 3, controlling for all attributes as before; the results are robust to employing alternative model specifications. Base levels are estimated, even though not reported due to space restrictions. Additional detailed results are presented in a web appendix:
http://people.stern.nyu.edu/padamopo/Adamopoulos_2021_MISQ_web_appendix.pdf

<div align="center">(Insert Table 5 about here)</div>

Moreover, we examine the validity of the social proof mechanism from the persuasion theory, too, in several additional ways. If the corresponding identified mechanism is valid, then the popularity of the recommended alternatives should have a positive and significant moderating effect on the effectiveness of recommendations as such alternatives are benefiting more from social proofs. Alternatively, if this identified mechanism is not valid, then more popular alternatives should have a negative or insignificant moderating effect. Thus, we investigate the effect of the different recommendation strategies across different levels of venue popularity, using the number of visits as a metric of popularity and employing the technique of quantile regression. Based on the results presented in Table 6, *recommendations have a stronger positive effect on average for more popular alternatives*. Hence, this further validates the persuasive role of recommendation strategies that embed social proofs. An interesting observation is that *the effect of quality and expert recommendations is much stronger for more popular alternatives,* *whereas the effect of trending recommendations is still positive but more stable across alternatives*. Prior research on desktop RSs had not identified such differences suggesting that "differences across recommender technologies are rather small" (Matt et al. 2013). The results are very similar to examining the moderating effect of the *number of reviews*, further verifying the identified mechanism of social proofs.

(Insert Table 6 about here)

Furthermore, we examine certain moderating effects of contextual factors and user attributes to further verify the theoretical foundations of the discussed mechanisms. In particular, according to the persuasion theory, consumers in positive mood are prone to persuasion techniques and exhibit increased reliance on cues and judgement heuristics (Mackie and Worth 1989). Hence, contextual factors might moderate the effectiveness of recommendations in the mobile channel as information processing in mobile devices is affect-driven and the distinctiveness of recommended alternatives would be further heightened. Thus, we hypothesize that contextual factors that affect people's mood states moderate the effectiveness of recommendations, as the effectiveness of persuasion mechanisms depends on mood states. Given the context of usage of mobile devices and the increased impact of the external environment, one important contextual factor that affects consumers' moods is the road traffic. Table 7 presents *the negative and significant effects of traffic levels on the effectiveness of recommendation strategies in the mobile channel*; the local traffic is measured based on the speed (miles per hour) of bike trips towards the venue based on data from the Divvy Bikes system in the city of Chicago.

To further verify the identified effects and their theoretical foundations, we then examine the moderating effect of additional contextual variables that directly affect mobile users' mood. More specifically, weather conditions, including exposure to sunlight and temperature levels, have been found to significantly affect consumers' moods as explained by certain biological processes (Schwarz and Clore 1983). Similarly, holidays and leisure time improve consumers' mood and promote well-being. Moreover, in the mobile context in particular, the reliability of the mobile network, the speed of the network, the latency and ubiquity of the mobile data connection and other aspects of the mobile network connection can induce frustration to users altering their moods and influencing their cognitive decision-making processes in the case of affect-driven processing; the overall mobile network performance is measured by Root Wireless, Inc. for the markets and time periods in our dataset. Thus, we study the moderating effect of the contextual factors of weather conditions, holidays, and the quality of the mobile network. Based on the results, holidays, better weather conditions, and increased mobile network performance have a positive and significant effect on the effectiveness of recommendations further illustrating that contextual factors, including conditions that relate to consumers' mood, feelings and affect, moderate the effectiveness of recommendations in the mobile channel as they determine whether consumers are likely to follow the provided recommendations. It is also worth noting that these moderating effects are stronger for recommendation strategies characterized by significant temporal diversity. This interesting finding finds support in prior works indicating that consumers at a state of high affect are more prone to variety-seeking behaviors (Li et al. 2017).

(Insert Table 7 about here)

Similarly, we examine the moderating effect of income on the effectiveness of recommendation strategies to further assess the theoretical foundations of the identified mechanism. In particular, one would expect that the effect of recommendations would be more pronounced for users with higher income as search costs are relatively higher for those users. However, as mobile devices are associated with affect-driven information processing and discretionary consumption, we hypothesize that recommendations are more effective for users with lower income as they might experience higher guilt for monetary expenses. That is, recommendations may provide to lower income consumers the necessary persuasion mechanism to construct reasons for justification of indulgence to monetary expenses reducing the associated guilt and making recommended alternatives much easier to choose (Shafir et al. 1993). Table 8 shows *the negative and significant effect of income on the effectiveness of recommendation strategies in the mobile channel*; user income is measured in USD (in 1000s) at the local zip code based on the American Community Survey 5-year estimates of the US Census Bureau. This finding further confirms the theoretical foundations of the mechanisms illustrating that recommendations provide the necessary persuasion mechanism to consumers for justifying and explaining their choices, mitigating any negative attributions and feelings associated with monetary expenses while making recommended alternatives much easier to choose.

(Insert Table 8 about here)

## 7. Robustness Checks

We conduct various robustness checks including, among others, instrumental variable techniques, alternative econometric model specifications, falsification tests, out-of-sample validation, and a replication study, as described in the next paragraphs.

### 7.1 Instrumental Variables for Endogenous Recommendation Strategies

We first employ instrumental variable techniques to further control for potential endogeneity in recommendations. Table 9 presents the heterogeneous effects of the different mobile recommendation strategies using the novel instruments we developed based on the employed deep-learning techniques (see Section 5 for a detailed description) and variables used in generating the recommendations to account for potential endogeneity in recommendation strategies. These novel instrumental variables, corresponding to the average and standard deviation of the alternative differentiation (isolation) for the specific time period, satisfy the relevance and exclusion criteria as they are related to whether the mobile application recommends an alternative (i.e., alternatives that are more isolated in the product space have a higher likelihood of being included in a recommendation list because of the diversification process of the employed recommendation algorithms) while they are determined prior to the revelation of $\xi_{jrt}$ (Nevo 2000). In addition, the instruments also include lags of the standardized percentage change in the number of photos and positive ratings, as well as the lag of the within-category standardized rating and number of photos. Hence, all the employed instruments are specific variables used in the algorithms to generate the recommendations but do not directly affect the utility of the alternatives for consumers in the exact current time period given the observed confounders in our econometric specifications; note that the sentiment of the reviews about each alternative is already included in the econometric specifications as measured directly by the mobile app as well as that the instruments are affected by other alternatives in the same market and that we do not rely on product descriptions written by managers.

We have tested the *validity of the instruments* based on several statistical tests. In particular, we have tested for under-identification of the models using the Kleibergen-Paap rank statistic concluding that the models are identified, and for weak identification using the Cragg-Donald statistic and the Stock-Yogo critical values in addition to the Kleibergen-Paap rank statistic, as well as for overidentification using the Sargan-Hansen test and Hansen's statistic. Based on all test statistics, we have confirmed the validity of the instruments.

As shown in Table 9, the results corroborate our previous findings regarding the impact of the various recommendation strategies. We should note that all models include fixed effects as well as time-varying controls for the venue, climate, geospatial, and calendar attributes, even though not depicted in the table due to space restrictions.

(Insert Table 9 about here)

We further test the *robustness of the instrumental variable models* based on domain adaptation for deep learning. The results are very similar after updating the word vectors by training the machine-learning model on the corpus of the mobile application itself. That is, instead of "freezing" the weights of the word layer as before when the pre-trained model was employed, we now render the word layer trainable and update during the training phase these weights, too, and not only the weights of the entity layer. The empirical results also remain the same when treating each individual user-generated review as a single document in our corpus with the entity layer of the neural model corresponding to the alternatives. Besides, the results remain robust to employing alternative deep-learning methods for latent space representations. For these alternative representations, we employ deep bidirectional latent feature-based contextual representations based on Devlin et al. (2018).[10] Similarly, the results remain robust to excluding from the instruments any latent dimensions of the representation that are correlated with any covariates in our empirical models, further alleviating any concerns for causal inference even if other confounders too are endogenous.

### 7.2 Additional Instrumental Variable Applications and Robustness Checks

We then employ instrumental variable techniques to further control for potential endogeneity in prices and within-group shares. In particular, Table 10 uses rental prices and the average price of other alternatives in the same market and same category and rating as well as the novel metric of alternative differentiation based on the employed machine-learning model to account for potential endogeneity in prices and within-group share. The motivation for the latter econometric instruments is similar to the seminal BLP-style instruments (Berry et al. 1995; Berry 1994), which measure the isolation in the observable product-characteristics space as products that are more isolated in a market are related to higher margins. There are also certain advantages of the developed instruments –compared to traditional instruments– in our setting worth noting as they further facilitate the identification in our study: the instruments would not directly enter the utility equation and they are not directly determined by managers. It is worth noting that other potential BLP-style instruments in the observable space would not be available in our setting as various observable characteristics might be determined by managers or might not be time-varying while most alternatives might not be available in other markets. Besides, we have tested the instruments' validity using the same tests and statistics as in Section 7.1. As shown in Table 10, the results further corroborate our findings.

---

[10] As another robustness check, we developed econometric instruments using a bag-of-words model (i.e., TF-IDF representation) but these instruments did not survive the validity tests. Such traditional approaches not surviving the validity tests of econometric instrumental variables finds theoretical support on the observation that they work in terms of discrete tokens without meaning while suffering from data sparsity and overfitting issues. This observation and comparison against traditional approaches further supports the proposed approach for instrumental variables.

### 7.3 Additional Robustness Checks

We conduct additional checks to ensure the robustness and generalizability of the findings; detailed results are presented in the web appendix.

One might be concerned, for instance, that the popularity of the mobile app might drive the results regarding the effectiveness of different recommendation strategies. We evaluate this possibility by capturing the effect of market-specific web search trends regarding the specific mobile app using data from Google Trends. The results remain highly robust.

Moreover, one might be concerned that there are alternative sources of information for users driving the results. Hence, we further supplement our dataset with the venue rating from Yelp.com, a major alternative source of information for users. The results remain robust.

Besides, one might also be concerned that there might be time-varying physical constraints driving the results. Hence, for each venue in our dataset, we include physical constraints based on disruptions of public transportation, local traffic, and (non-)availability of different means of transportation, based on open data from the Chicago Transit Authority and the Divvy Bikes local sharing system. The results remain highly robust, further corroborating our findings.

Additionally, we control for the lagged effect of recommendations in our econometric specifications. Based on the results, the lagged effect of the recommendations captures a small portion of the previously identified effect as the estimated coefficients slightly decreased; however, the demand effects of recommendations are not driven by just the past recommendations. This is an interesting finding further verifying the underlying mechanism as – in accordance with the systemic-heuristic theory of persuasion– the social proof mechanism tends to be less persistent (Bhattacherjee and Sanford 2006).

In addition, we check the robustness of our findings using several alternative model specifications, including seasonality effects and category-specific non-linear time trends. All the results remain robust, further corroborating our findings.

We further incorporate and account for heterogeneity across additional dimensions by assigning random coefficients to the corresponding effects in the econometric specifications. For instance, we consider the heterogeneity across implementation details of the various recommendation strategies by taking advantage of the different (software) releases of the mobile app over the time period of our panel. The results corroborate our findings. The results also remain robust after incorporating heterogeneity across implementation details based on merges of development code to production. In addition, we further incorporate heterogeneity by assigning random coefficients across markets to the main variables of interest. The results corroborate our previous findings.

Furthermore, in order to enhance the homogeneity of our dataset, we conduct a subsample analysis and we estimate the effectiveness of the various recommendation strategies only for alternatives that have been recommended before. The results corroborate our previous findings. In addition, we test for homogeneous non-causality (Dumitrescu and Hurlin 2012) and –based on the Wald statistics– we reject the null hypothesis that no causal relationship exists between the dependent variable and the variables of interest.

### 7.3.2 Falsification Tests

We additionally run different falsification tests ("placebo" studies) further verifying that the previous findings are not a statistical artifact, but that we indeed discovered the actual effects. In particular, we run different "placebo" studies using the same models as above (to maintain consistency) but randomly indicating i) which alternatives (i.e., random alternative recommended) were recommended and ii) when (i.e., random time period of recommendation), respectively. Under these checks, the corresponding effects are not statistically significant, indicating that our previous findings are not a statistical artifact of our specifications, but that we indeed discovered the actual effects.

### 7.3.3 Out-of-sample Validation

Moreover, we further validate the findings assessing the out-of-sample performance of the empirical models and verifying that they generalize well beyond in-sample observations. In particular, we employ a hold-out evaluation scheme with an 80/20 chronological split of the data (i.e., the latest time periods are used for evaluation) and evaluate each model based on the metrics of root-mean-square error (RMSE), mean-square error (MSE), mean absolute deviation (MAD), and mean absolute percent error (MAPE). Based on Tables A3 and A4 in the appendix, all the employed models and econometric specifications exhibit very good out-of-sample performance; a baseline method of predicting the average demand level for each alternative leads to an out-of-sample RMSE of 1.3199. These results also illustrate that the reported explanatory power is not due to over-fitting the data, but we indeed unveil the effect of various recommendation strategies, and our findings generalize well beyond in-sample observations.

### 7.3.4 Replication Study
Finally, we further empirically validate the generalizability of the results by conducting a replication study repeating the analyses using the same RS strategies but in the domain of nightlife entertainment. All the results are robust and further corroborate our previous findings.

Overall, we have examined the robustness of the results in a variety of ways and we have illustrated that the findings of this study are highly robust to various alternative econometric specifications, robustness checks and falsification tests, as well as that they extend beyond the main empirical setting we examined.

## 8. Discussion and Managerial Implications
The findings of this study, apart from a theoretical contribution, have important managerial implications. Our analyses reveal that recommendations in the mobile context have a *positive* and economically significant effect on individual demand levels for the recommended alternatives but the effects are largely *heterogeneous* across popular recommendation strategies. This is an important and timely finding for managers as mobile recommender systems currently have been adopted by only 13% of the companies across the globe, even though RSs are the most widespread personalization technique for traditional shopping channels (Econsultancy.com 2013). It also highlights that RSs can play an important role in the business-expanding strategies of firms in the mobile channel. Apart from unveiling the economic impact of recommendations in the mobile context and their effects on consumers' decision-making, we illustrate how managers and practitioners can leverage observational data to evaluate RS algorithms and estimate their corresponding causal demand effects. This provides a valuable and non-intrusive tool for managers to analyze and compare the performance of their recommendation strategies for their businesses. This is especially important nowadays as more than half of the companies (57%) do not test the performance of their own RSs (Econsultancy.com 2013; Henrion 2019). The structural econometric model and the novel machine-learning instruments allow for actual demand estimation, without incurring any additional costs. Hence, beyond estimating the heterogeneous effects, the proposed method provides accurate predictions of the impact on individual demand in cases of entries and exits of alternatives and the design of optimal features for new alternatives better informing firm strategies.

Moreover, disentangling the effects and economic impact of various types of mobile recommendations, we find that the various recommendation strategies exhibit significant heterogeneity in the mobile context. Interestingly, recommendation strategies that provide "in-the-moment" content to users have a much stronger effect compared to other commonly used recommendations based on just historical trends and data. Such findings are essential for managers and practitioners alike as they can guide the informed adoption of recommendation strategies that are particularly effective in the mobile channel. Similarly, they might help businesses better adjust and finetune or market their recommendation strategies. For instance, in cases of hybrid recommendations, managers may emphasize and communicate the social proof components of their recommendations. The importance and economic significance of these findings is further amplified by the prediction that in the coming years, the number of mobile shoppers will reach 223.7 million, with 83% of smartphone users shopping online using their mobile devices (eMarketer 2019). Besides, such findings can better inform the development of more effective RSs in the future. For instance, managers and practitioners can leverage our findings to design novel RS algorithms that are characterized by significant temporal diversity and embed various social proofs in order to further enhance product sales combining the persuasive and informative aspects of recommendations. In addition, the findings highlight the incremental value of capturing and leveraging "in-the-moment" data in a mobile context. Such data could be leveraged by firms to enhance the effectiveness of other strategic tools too.

Similarly, examining the moderating effects of various item attributes yields valuable insights into consumer behavior and a much-needed understanding of the heterogeneity of the effectiveness of various recommendation strategies in the mobile setting. Such moderating effects too are substantial for businesses and RS practitioners as they highlight the importance of various product-design decisions managers should consider in order to improve the economic performance of their RSs. In addition, managers can further understand which alternatives would accrue the highest benefit if recommended to consumers and which would benefit the least. For instance, the moderating effect of price on the effectiveness of recommendation strategies showcases that even though more expensive alternatives are less appealing to the consumers, they can more effectively leverage the additional attention they garner from a recommendation, suggesting that certain RS strategies lend themselves well to upselling opportunities. Such findings also alleviate certain managerial concerns that recommendation strategies might be pushing consumers towards items with lower margins hurting the profitability of firms despite the elevated product demand levels. Hence, mobile RSs not only increase total demand levels but also create business advantage by inducing upselling opportunities. Another actionable finding of significant importance for managers is that novel alternatives accrue greater benefits from recommendations when those recommendations are not based solely on the characteristic of novelty but also take into consideration specific item attributes, such as the quality of the item. Such findings can help businesses avoid any "blind spots" by selecting the right recommendation strategies to maximize demand levels for their own product mixture. Besides, yet another important finding with significant managerial implications is that the effect of several common recommendation strategies is much stronger for more popular products, whereas the effect of trending recommendations is more stable across alternatives, allowing businesses to effectively leverage the benefits of long-tail products. Such findings also enable managers to better understand fluctuation in demand and better manage the demand levels of each alternative, avoiding unintended consequences. Additionally, these findings highlight the capability of certain recommendation strategies to avoid a winner-takes-all market. Overall, the managerial importance of these findings and the detailed understanding of the effectiveness of RSs is further highlighted by the fact that 78% of the companies consider lack of knowledge as a barrier to adopting or improving RSs in their organization and 83% of marketers consider this their biggest challenge (Econsultancy.com 2013; Monetate 2019).

In addition, the moderating effects of the context on the effectiveness of different recommendation strategies, too, have intriguing implications for businesses. For instance, the identified differences across contextual factors suggest that practitioners can leverage this source of heterogeneity to design more effective content ranking approaches, taking advantage of such context dynamics. Such a ranking approach could directly incorporate contextual factors and dynamically rank various recommendation strategies or different information-providing mechanisms in order to maximize the effectiveness for each context directly. Hence, RSs can tap into the unique aspects of mobile settings to further enhance their effectiveness. This is an implication of significant potential as there is a lack of rigorous understanding of how marketplaces and platforms should select recommendation candidates under various contextual factors. For instance, items related to more discretionary and hedonic consumption or more diverse items related to variety-seeking may be recommended at cases of more positive affect, while recommending more essential options in cases of neutral or negative affect. Yet another implication of significant managerial interest is that brands and managers can use these findings to select each time the most effective context for the delivery of recommendations or promotional messages. For instance, our findings illustrate that recommendation strategies are more effective when traffic levels are lower or mobile networks perform better. Given the various costs of firm-consumer communications and the likelihood of inducing consumer annoyance (Todri et al. 2020), managers can use these findings to avoid wasting opportunities for recommendations or other promotions in contexts associated with lower effectiveness as the marginal cost of delivering recommendations might outweigh their marginal benefit for certain alternatives in certain settings due to the corresponding heterogeneity.

Furthermore, the moderating effects of user characteristics on the effectiveness of the different recommendation strategies also have direct implications for managers. For instance, managers may adjust their mobile applications to highlight different item attributes or recommendation strategies to different users as the relative importance of these characteristics is systematically changing across demographics, for example.

There are several other actionable managerial implications based on the conducted analyses. For instance, incorporating several sources of heterogeneity, managers and practitioners have additional information regarding how the demand effects of the various recommendation strategies might vary across local markets, domains, implementations, users, etc. allowing for more accurate demand predictions further improving the efficacy of operations and resource planning, among other managerial activities.

Future research, apart from estimating the impact of recommender systems across additional settings in the mobile context, can apply the proposed techniques in order to generate contextual recommendation lists based on consumers' surplus. Similarly, future research can consider explicit user satisfaction, instead of demand effects. Future research may also compare the effectiveness of mobile recommendations vis-à-vis recommendations on other channels.

# Tables and Figures

| Table 1: Descriptive Statistics | | | | |
|---|---|---|---|---|
| **Variable** | **Mean** | **Std. Dev.** | **Min** | **Max** |
| Visitor share | 0.000 | 0.001 | 0.000 | 0.009 |
| Trending recommendation | 0.008 | 0.091 | 0.000 | 1.000 |
| Quality recommendation | 0.008 | 0.075 | 0.000 | 1.000 |
| Event recommendation | 0.004 | 0.038 | 0.000 | 1.000 |
| Expert recommendation | 0.006 | 0.050 | 0.000 | 1.000 |
| Novel recommendation | 0.002 | 0.034 | 0.000 | 1.000 |
| Price | 1.791 | 0.744 | 1.000 | 4.000 |
| Rating | 3.321 | 1.385 | 0.000 | 5.000 |
| Number of Reviews | 44.138 | 56.380 | 1.000 | 1,088.000 |
| Sentiment of Reviews | 0.880 | 0.380 | -0.997 | 1.000 |
| Photos | 134.766 | 234.706 | 1.000 | 5,660.000 |
| Chain | 0.456 | 0.498 | 0.000 | 1.000 |
| Marketing promotions | 0.060 | 0.238 | 0.000 | 1.000 |
| Alcohol | 0.555 | 0.497 | 0.000 | 1.000 |
| Delivery | 0.387 | 0.487 | 0.000 | 1.000 |
| Takeout | 0.847 | 0.360 | 0.000 | 1.000 |
| Reservations | 0.466 | 0.499 | 0.000 | 1.000 |
| Credit cards | 0.976 | 0.153 | 0.000 | 1.000 |
| Outdoor seating | 0.421 | 0.494 | 0.000 | 1.000 |
| Wi-Fi | 0.280 | 0.449 | 0.000 | 1.000 |
| Parking | 0.005 | 0.069 | 0.000 | 1.000 |
| Wheelchair accessible | 0.023 | 0.150 | 0.000 | 1.000 |
| TVs | 0.009 | 0.096 | 0.000 | 1.000 |
| Music | 0.012 | 0.111 | 0.000 | 1.000 |
| Holidays | 0.072 | 0.258 | 0.000 | 1.000 |
| Precipitation | 0.086 | 0.267 | 0.000 | 4.240 |
| Temperature | 55.237 | 12.872 | 10.000 | 81.000 |
| Mobile network performance | 1.537 | 0.708 | 1.000 | 3.000 |
| Traffic | 6.166 | 1.468 | 0.000 | 11.844 |
| Income | 44257.08 | 19538.04 | 5377.00 | 168839.00 |

| Table 2: Employed Mathematical Notation | |
|---|---|
| **Symbol** | **Description** |
| $r \in \{1, \ldots, R\}$ | Markets (cities) |
| $j \in \{0, 1, \ldots, N\}$ | Alternative (venue). Note that $j = 0$ represents the "outside option", which corresponds to alternatives that might not be present in our data set or the option of a user not visiting any venue at all |
| $t \in \{1, \ldots, T\}$ | Time period (day) |
| $s_{jrt}$ | Market share of alternative $j$ in marker $r$ in (time) period $t$ |
| $\bar{s}_{j/g}$ | Market share of alternative $j$ as a fraction of the total group (nest) share $g$ |
| $p_{jrt}$ | Price of alternative $j$ in marker $r$ in time period $t$ |
| $z_{jrt}$ | Observed alternative characteristics (e.g., rating, number of reviews, photos) |
| $w_{jrt}$ | Contextual factors (e.g., temperature, holidays) |
| $\rho_{jrt}$ | Recommendation types (e.g., event recommendation) |
| $\xi_j$ | Unobserved characteristics |
| $\varepsilon_{ij}$ | User-specific taste parameters and factors quixotic to the consumer himself |
| $\sigma$ | With-in group (nest) correlation |

| Table 3: Coefficient Estimates of Nested Logit Demand Model | | | |
|---|---|---|---|
| | **Model 1** | **Model 2** | **Model 3** |
| Recommendation | 0.9712*** | | |
| Trending recommendation | | 1.4935*** | 1.4908*** |
| Quality recommendation | | 0.9436*** | 1.0134*** |
| Event recommendation | | 0.2687*** | 0.3562*** |
| Expert recommendation | | 1.0730*** | 1.1444*** |
| Novel recommendation | | 0.2535*** | 0.3399*** |
| Recommendation ranking | | | -0.0019*** |
| Price | -0.0177*** | -0.0164*** | -0.0163*** |
| Rating | 0.0030* | 0.0050*** | 0.0039** |
| Number of reviews (log) | 0.1627*** | 0.1543*** | 0.1532*** |
| Sentiment of reviews | 0.0213*** | 0.0203*** | 0.0203*** |
| Photos (log) | 0.0510*** | 0.0515*** | 0.0522*** |
| Chain | 0.0413*** | 0.0392*** | 0.0397*** |
| Marketing promotions | -0.0033*** | -0.0032*** | -0.0032*** |
| Alcohol | 0.0509*** | 0.0516*** | 0.0520*** |
| Delivery | -0.0575*** | -0.0563*** | -0.0562*** |
| Takeout | -0.0406*** | -0.0492*** | -0.0493*** |
| Reservations | 0.0198*** | 0.0182*** | 0.0184*** |
| Credit cards | 0.0552*** | 0.0535*** | 0.0535*** |
| Outdoor seating | -0.0047 | -0.0024 | -0.0023 |
| Wi-Fi | 0.0149*** | 0.0167*** | 0.0166*** |
| Parking | 0.0759*** | 0.0687*** | 0.0657*** |
| Wheelchair accessible | 0.0116 | 0.0055 | 0.0057 |
| TVs | 0.0116 | 0.0121 | 0.0124 |
| Music | 0.0341*** | 0.0338*** | 0.0343*** |
| Within-group share | 0.1127*** | 0.1114*** | 0.1111*** |
| Market-level fixed effects | Yes | Yes | Yes |
| Category-level fixed effects | Yes | Yes | Yes |
| Additional alternative controls | Yes | Yes | Yes |
| Context controls | Yes | Yes | Yes |
| Time trend | Yes | Yes | Yes |
| Log-likelihood | -843649 | -841686 | -841639 |
| Adjusted R-squared | 0.541 | 0.543 | 0.543 |
| *N* | 711,673 | 711,673 | 711,673 |

Note: Panel data analysis of nested logit demand model with multi-level fixed effects corresponding to the market and the alternative category. The estimation sample includes observations about all available alternatives in each market and time period. The number of reviews variable corresponds to the log of the number of user-generated reviews for each alternative and the photos variable corresponds to the log of the photos posted by users for each alternative. The additional alternative controls include the hours of operation of the specific alternative, what meals are served, for how many weeks (log) the alternative has been operating, whether the venue is part of a chain, and the number of events in the specific alternative and time period. The context controls include day-of-the-week and holiday effects, local temperature and precipitation levels, as well as the average geographic distance of alternative. The specifications also include a linear time trend. Significance levels: * p<0.05, ** p<0.01, *** p<0.001

| Table 4: Moderating Effects – Price | | | |
|---|---|---|---|
| | **Model 1** | **Model 2** | **Model 3** |
| Recommendation x Price | 0.0515*** | | |
| Trending recommendation x Price | | 0.0409* | 0.0419* |
| Quality recommendation x Price | | 0.0130 | 0.0132 |
| Event recommendation x Price | | -0.1705*** | -0.1835*** |
| Expert recommendation x Price | | 0.0666 | 0.0863* |
| Novel recommendation x Price | | 0.1967*** | 0.2044*** |
| Market-level fixed effects | Yes | Yes | Yes |
| Category-level fixed effects | Yes | Yes | Yes |
| Additional alternative controls | Yes | Yes | Yes |
| Context controls | Yes | Yes | Yes |
| Time trend | Yes | Yes | Yes |
| Log-likelihood | -843614 | -841663 | -841601 |
| Adjusted R-squared | 0.541 | 0.543 | 0.543 |
| *N* | 711,673 | 711,673 | 711,673 |

Note: Panel data analysis of nested logit demand model with multi-level fixed effects corresponding to the market and the alternative category. Additional notes as in Table 1. Significance levels: * p<0.05, ** p<0.01, *** p<0.001

| Table 5: Moderating Effects – Novelty | | | |
|---|---|---|---|
| | **Model 1** | **Model 2** | **Model 3** |
| Recommendation x Weeks open | -0.1060*** | | |
| Trending recommendation x Weeks open | | -0.2230*** | -0.2226*** |
| Quality recommendation x Weeks open | | -0.2000*** | -0.1992*** |
| Event recommendation x Weeks open | | 0.0794* | 0.0645 |
| Expert recommendation x Weeks open | | 0.0393 | 0.0358 |
| Novel recommendation x Weeks open | | 0.0750*** | 0.0747*** |
| Market-level fixed effects | Yes | Yes | Yes |
| Category-level fixed effects | Yes | Yes | Yes |
| Additional alternative controls | Yes | Yes | Yes |
| Context controls | Yes | Yes | Yes |
| Time trend | Yes | Yes | Yes |
| Log-likelihood | -843446 | -841190 | -841133 |
| Adjusted R-squared | 0.541 | 0.544 | 0.544 |

| | | N | | 711,673 | | 711,673 | | 711,673 |
|---|---|---|---|---|---|---|---|---|

Note: Panel data analysis of nested logit demand model with multi-level fixed effects corresponding to the market and the alternative category. Additional notes as in Table 1. Significance levels: * p<0.05, ** p<0.01, *** p<0.001

## Table 6: Coefficient Estimates of Nested Logit Demand Model - Quantile Regression

| | Q: 0.10 | Q: 0.20 | Q: 0.30 | Q: 0.40 | Q: 0.50 | Q: 0.60 | Q: 0.70 | Q: 0.80 | Q: 0.90 |
|---|---|---|---|---|---|---|---|---|---|
| Trending recommendation | 1.297*** | 1.830*** | 2.080*** | 2.240*** | 1.819*** | 1.632*** | 1.567*** | 1.475*** | 1.269*** |
| Quality recommendation | 0.015*** | 0.119*** | 0.487*** | 0.808*** | 0.671*** | 0.630*** | 0.702*** | 0.983*** | 4.244*** |
| Event recommendation | -0.016*** | -0.219*** | -0.500*** | -0.693*** | -0.442*** | -0.280*** | -0.172*** | 0.194*** | 3.649*** |
| Expert recommendation | 0.045*** | 0.337*** | 0.863*** | 1.009*** | 1.033*** | 1.040*** | 1.244*** | 1.629*** | 2.255*** |
| Novel recommendation | 0.012*** | 0.092*** | 0.429*** | 0.744*** | 0.599*** | 0.538*** | 0.530*** | 0.499*** | 1.085*** |
| Market-level fixed effects | Yes | Yes | Yes | Yes | Yes | Yes | Yes | Yes | Yes |
| Category-level fixed effects | Yes | Yes | Yes | Yes | Yes | Yes | Yes | Yes | Yes |
| Additional alternative controls | Yes | Yes | Yes | Yes | Yes | Yes | Yes | Yes | Yes |
| Context controls | Yes | Yes | Yes | Yes | Yes | Yes | Yes | Yes | Yes |
| Time trend | Yes | Yes | Yes | Yes | Yes | Yes | Yes | Yes | Yes |
| Pseudo R-squared | 0.505 | 0.508 | 0.508 | 0.491 | 0.456 | 0.453 | 0.441 | 0.419 | 0.409 |
| N | 711,673 | 711,673 | 711,673 | 711,673 | 711,673 | 711,673 | 711,673 | 711,673 | 711,673 |

Note: Quantile regression analysis of nested logit demand model with multi-level fixed effects corresponding to the market and the alternative category. Additional notes as in Table 1. Significance levels: * p<0.05, ** p<0.01, *** p<0.001

## Table 7: Moderating Effects – Traffic

| | Model 1 | Model 2 | Model 3 |
|---|---|---|---|
| Recommendation x Traffic | -0.0580*** | | |
| Trending recommendation x Traffic | | -0.1963*** | -0.1961*** |
| Quality recommendation x Traffic | | -0.0638*** | -0.0644*** |
| Event recommendation x Traffic | | -0.2223*** | -0.2196*** |
| Expert recommendation x Traffic | | -0.1428** | -0.1418** |
| Novel recommendation x Traffic | | 0.0525 | 0.0518 |
| Market-level fixed effects | Yes | Yes | Yes |
| Category-level fixed effects | Yes | Yes | Yes |
| Additional alternative controls | Yes | Yes | Yes |
| Context controls | Yes | Yes | Yes |
| Time trend | Yes | Yes | Yes |
| Log-likelihood | -108670 | -108275 | -108271 |
| Adjusted R-squared | 0.296 | 0.302 | 0.302 |
| p | 0.0000 | 0.0000 | 0.0000 |
| N | 93,482 | 93,482 | 93,482 |

Note: Panel data analysis of nested logit demand model with multi-level fixed effects corresponding to the market and the alternative category. The local traffic is measured based on the speed (miles per hour) of bike trips to the venue based on open data from the Divvy Bikes local sharing system in the city of Chicago. Additional notes as in Table 1. Significance levels: * p<0.05, ** p<0.01, *** p<0.001

## Table 8: Moderating Effects – Income

| | Model 1 | Model 2 | Model 3 |
|---|---|---|---|

| | Model 1 | Model 2 | Model 3 |
|---|---|---|---|
| Recommendation x Income | -0.0015*** | | |
| Trending recommendation x Income | | -0.0037*** | -0.0037*** |
| Quality recommendation x Income | | -0.0034*** | -0.0030*** |
| Event recommendation x Income | | 0.0110*** | 0.0129*** |
| Expert recommendation x Income | | -0.0078*** | -0.0078*** |
| Novel recommendation x Income | | -0.0071** | -0.0063* |
| Market-level fixed effects | Yes | Yes | Yes |
| Category-level fixed effects | Yes | Yes | Yes |
| Additional alternative controls | Yes | Yes | Yes |
| Context controls | Yes | Yes | Yes |
| User controls | Yes | Yes | Yes |
| Time trend | Yes | Yes | Yes |
| Log-likelihood | -843574 | -841664 | -841603 |
| Adjusted R-squared | 0.541 | 0.543 | 0.543 |
| *N* | 711,673 | 711,673 | 711,673 |

Note: Panel data analysis of nested logit demand model with multi-level fixed effects corresponding to the market and the alternative category. The user controls include the user income. The user income is measured in USD (in 1000s) at the local zip code of the venue based on the American Community Survey 5-year estimates of the US Census Bureau. Additional notes as in Table 1. Significance levels: * p<0.05, ** p<0.01, *** p<0.001

| Table 9: Coefficient Estimates of Nested Logit Demand Model with Instrumental Variables | | | |
|---|---|---|---|
| | **Model 1** | **Model 2** | **Model 3** |
| Recommendation | 1.1861*** | | |
| Trending recommendation | | 1.4480*** | 1.5122*** |
| Quality recommendation | | 0.8638*** | 1.0244*** |
| Event recommendation | | 0.3501*** | 0.3810*** |
| Expert recommendation | | 1.0934*** | 1.1277*** |
| Novel recommendation | | 0.2736*** | 0.3754*** |
| Recommendation ranking | | | -0.0021*** |
| Price | -0.0163*** | -0.0164*** | -0.0151*** |
| Rating | 0.0024 | 0.0073*** | -0.0004 |
| Number of reviews (log) | 0.1572*** | 0.1580*** | 0.1494*** |
| Sentiment of reviews | 0.0201*** | 0.0204*** | 0.0204*** |
| Photos (log) | 0.0446*** | 0.0528*** | 0.0520*** |
| Marketing promotions | -0.0029*** | -0.0032*** | -0.0029*** |
| Within-group share | 0.1140** | 0.1138*** | 0.1134*** |
| Market-level fixed effects | Yes | Yes | Yes |
| Category-level fixed effects | Yes | Yes | Yes |
| Additional alternative controls | Yes | Yes | Yes |
| Context controls | Yes | Yes | Yes |
| Time trend | Yes | Yes | Yes |
| Log-likelihood | -788745 | -786170 | -788587 |
| Adjusted R-squared | 0.547 | 0.551 | 0.547 |
| *N* | 680,404 | 680,404 | 680,404 |

Note: Panel data analysis of nested logit demand model with instrumental variables. The instrumental variables for recommendation strategies include the average and standard deviation of the alternative differentiation and isolation based on the employed machine-learning model of the user-generated reviews, lags of the standardized percentage change in the number of photos and positive ratings, as well as the lag of the within-category standardized rating and number of photos. Additional notes as in Table 1. Significance levels: * p<0.05, ** p<0.01, *** p<0.001

| Table 10: Coefficient Estimates of Nested Logit Demand Model with Instrumental Variables | | | |
|---|---|---|---|
| | **Model 1** | **Model 2** | **Model 3** |
| Recommendation | 0.9780*** | | |
| Trending recommendation | | 1.5141*** | 1.5122*** |
| Quality recommendation | | 0.9472*** | 1.0244*** |
| Event recommendation | | 0.2828*** | 0.3810*** |
| Expert recommendation | | 1.0504*** | 1.1277*** |
| Novel recommendation | | 0.2800*** | 0.3754*** |
| Recommendation ranking | | | -0.0021*** |
| Price | -0.0438*** | -0.0407*** | -0.0398*** |
| Rating | 0.0043* | 0.0061** | 0.0049* |
| Number of reviews (log) | 0.1718*** | 0.1606*** | 0.1597*** |
| Sentiment of reviews | 0.0172*** | 0.0169*** | 0.0168*** |
| Photos (log) | 0.0552*** | 0.0553*** | 0.0560*** |
| Marketing promotions | -0.0033*** | -0.0032*** | -0.0032*** |
| Within-group share | 0.0892** | 0.0957*** | 0.0945*** |
| Market-level fixed effects | Yes | Yes | Yes |
| Category-level fixed effects | Yes | Yes | Yes |
| Additional alternative controls | Yes | Yes | Yes |
| Context controls | Yes | Yes | Yes |
| Time trend | Yes | Yes | Yes |
| Log-likelihood | -827001 | -824543 | -824536 |
| Adjusted R-squared | 0.542 | 0.546 | 0.546 |
| *N* | 700,478 | 700,478 | 700,478 |

Note: Panel data analysis of nested logit demand model with instrumental variables. The instrumental variables for price and within-group share include rental prices and the average price of other alternatives in the same market and same category and rating as well as the novel metric of alternative differentiation and isolation based on the employed machine-learning model of the user-generated reviews. Additional notes as in Table 1. Significance levels: * p<0.05, ** p<0.01, *** p<0.001

# Appendix

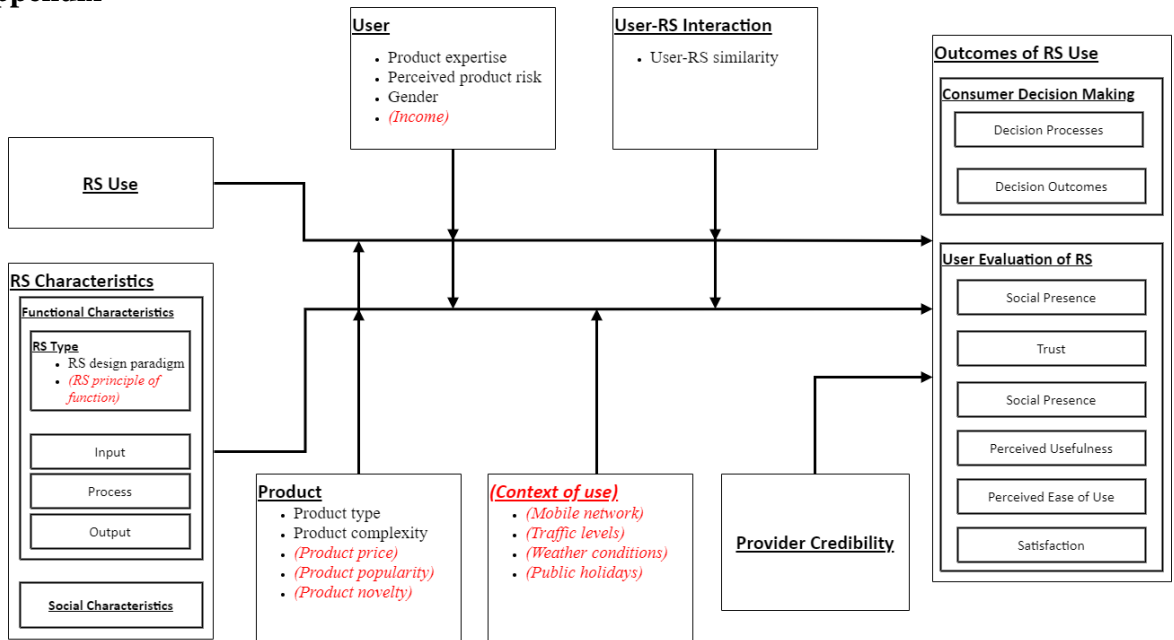

**Figure A1. Conceptual model of the effects of recommender systems.**

| Table A1: Coefficient Estimates of Nested Logit Demand Model with Fixed Effects | | | |
|---|---|---|---|
| | **Model 1** | **Model 2** | **Model 3** |
| Recommendation | 0.8893*** | | |
| Trending recommendation | | 1.0610*** | 1.0572*** |
| Quality recommendation | | 0.8903*** | 0.9765*** |
| Event recommendation | | 0.4896*** | 0.5877*** |
| Expert recommendation | | 0.8421*** | 0.9294*** |
| Novel recommendation | | -0.0629 | 0.0299 |
| Recommendation ranking | | | -0.0023*** |
| Rating | 0.0166*** | 0.0180*** | 0.0174*** |
| Number of reviews (log) | 0.1068*** | 0.1063*** | 0.1060*** |
| Sentiment of reviews | 0.0013 | 0.0012 | 0.0012 |
| Photos (log) | -0.0012 | -0.0012 | -0.0005 |
| Marketing promotions | 0.0024* | 0.0023* | 0.0023* |
| Within-group share | 0.2544*** | 0.2532*** | 0.2529*** |
| Alternative-level fixed effects | Yes | Yes | Yes |
| Additional alternative controls | Yes | Yes | Yes |
| Context controls | Yes | Yes | Yes |
| Time trend | Yes | Yes | Yes |
| Log-likelihood | -796329 | -795765 | -795667 |
| Adjusted R-squared | 0.592 | 0.593 | 0.598 |
| *p* | 0.0000 | 0.0000 | 0.0000 |
| *N* | 711,673 | 711,673 | 711,673 |

Note: Panel data analysis of nested logit demand model with alternative-level fixed effects. Additional notes as in Table 1. Significance levels: * p<0.05, ** p<0.01, *** p<0.001

| Table A2: Coefficient Estimates of Random Coefficients Demand Model | | | |
|---|---|---|---|
| | **Model 1** | **Model 2** | **Model 3** |
| Recommendation | 0.1268*** | | |
| Trending recommendation | | 1.4529*** | 1.4518*** |
| Quality recommendation | | 0.7181*** | 0.7815*** |
| Event recommendation | | 0.1476*** | 0.0477*** |
| Expert recommendation | | 1.2264*** | 1.2436*** |
| Novel recommendation | | 0.0118 | 0.0988*** |
| Recommendation ranking | | | -0.0015 |
| Price | -0.0089*** | -0.0086*** | -0.0088*** |
| Rating | 0.0295* | 0.0295*** | 0.0290*** |
| Number of reviews (log) | 0.2873*** | 0.2744*** | 0.2736*** |
| Sentiment of reviews | 0.0241*** | 0.0225 | 0.0224** |
| Photos (log) | 0.0156*** | 0.0182 | 0.0186 |
| Marketing promotions | -0.0017 | -0.0015 | -0.0015* |
| Within-group share | 0.1357*** | 0.1413*** | 0.1398*** |
| St. dev. Recommendation | 0.1893*** | | |
| St. dev. Trending recommendation | | 0.1650*** | 0.1076*** |
| St. dev. Quality recommendation | | 0.2380*** | 0.0099* |
| St. dev. Event recommendation | | 1.0618*** | 0.2351*** |
| St. dev. Expert recommendation | | 0.5683*** | 0.0899*** |
| St. dev. Novel recommendation | | 0.5764*** | 0.0213*** |
| Market-level fixed effects | Yes | Yes | Yes |
| Category-level fixed effects | Yes | Yes | Yes |
| Additional alternative controls | Yes | Yes | Yes |
| Context controls | Yes | Yes | Yes |
| Time trend | Yes | Yes | Yes |
| Objective value | 464.6449 | 461.7530 | 468.7254 |
| *p* | 0.0000 | 0.0000 | 0.0000 |
| *N* | 255,060 | 255,060 | 255,060 |

Note: Analysis of random coefficients nested logit model of demand (BLP). Complete panels were used for computational efficiency. Additional notes as in Table 1. Significance levels: * p<0.05, ** p<0.01, *** p<0.001

| Table A3: In-Sample Validation of Nested Logit Demand Model | | | |
|---|---|---|---|
| | **Model 1** | **Model 2** | **Model 3** |
| RMSE | 0.791763 | 0.789581 | 0.789514 |
| MSE | 0.626888 | 0.623439 | 0.623333 |
| MAD | 0.483424 | 0.480524 | 0.480507 |
| MAPE | 6.031686 | 6.005812 | 6.005087 |

Note: In-sample validation of panel data analysis of nested logit demand model. The sample corresponds to an 80/20 chronological split of the data. Additional notes as in Table 1.

| Table A4: Out-of-Sample Validation of Nested Logit Demand Model | | | |
|---|---|---|---|
| | **Model 1** | **Model 2** | **Model 3** |
| RMSE | 0.963149 | 0.957922 | 0.957720 |
| MSE | 0.927655 | 0.917615 | 0.917228 |
| MAD | 0.609633 | 0.602756 | 0.602656 |
| MAPE | 8.376500 | 8.323763 | 8.320875 |

Note: Out-of-sample validation of panel data analysis of nested logit demand model. Additional notes as in Table W28.